\documentclass[3p,times,procedia]{elsarticle}

\usepackage{ecrc}
\usepackage{algorithm}
\usepackage{algorithmic}
\usepackage{subfigure}
\usepackage{graphicx}
\usepackage[section] {placeins}
\usepackage{multirow}
\usepackage{amsmath}
\usepackage{amssymb}
\volume{00}
\firstpage{1}
\journalname{Procedia Computer Science}
\runauth{}
\jid{procs}
\jnltitlelogo{Procedia Computer Science}
\CopyrightLine{2013}{Published by Elsevier Ltd.}
\usepackage{amssymb}
\usepackage[figuresright]{rotating}
\begin{document}
\begin{frontmatter}
\dochead{International Workshop on Body Area Sensor Networks (BASNet-2013)}
\title{EDDEEC: Enhanced Developed Distributed Energy-Efficient\\ Clustering for Heterogeneous Wireless Sensor Networks}

\author{N. Javaid$^{\pounds}$, T. N. Qureshi$^{\pounds}$, A. H. Khan$^{\pounds}$, A. Iqbal$^{\pounds}$, E. Akhtar$^{\sharp}$, M. Ishfaq$^{\S}$}

\address{$^{\pounds}$COMSATS Institute of Information Technology, Islamabad, Pakistan. \\
        $^{\sharp}$University of Bedfordshire, Luton, UK.\\
        $^{\S}$King Abdulaziz University, Rabigh, Saudi Arabia.}

\begin{abstract}
Wireless Sensor Networks (WSNs) consist of large number of randomly deployed energy constrained sensor nodes. Sensor nodes have ability to sense and send sensed data to Base Station (BS). Sensing as well as transmitting data towards BS require high energy. In WSNs, saving energy and extending network lifetime are great challenges. Clustering is a key technique used to optimize energy consumption in WSNs. In this paper, we propose a novel clustering based routing technique: Enhanced Developed Distributed Energy Efficient Clustering scheme (EDDEEC) for heterogeneous WSNs. Our technique is based on changing dynamically and with more efficiency the Cluster Head (CH) election probability. Simulation results show that our proposed protocol achieves longer lifetime, stability period and more effective messages to BS than Distributed Energy Efficient Clustering (DEEC), Developed DEEC (DDEEC) and Enhanced DEEC (EDEEC) in heterogeneous environments.
\end{abstract}
\end{frontmatter}
\section{Background}
In WSNs, all the nodes have to send sensed data to BS, often called sink. Usually nodes in WSNs are power constrained due to limited battery resource. It is also not possible to recharge or replace the battery of already deployed sensor nodes \cite{talha,talha 1, Reference 1}.

Routing protocols play important role in achieving energy efficiency in WSNs. Clustering is used to minimize energy consumption. In this technique members of the cluster elect a CH \cite{Reference 2,Reference 3}. All nodes belonging to the same cluster send their data to CH, where, CH aggregates data and sends aggregated data to BS \cite{Reference 4,Reference 5,Reference 6}.

Clustering is useful in achieving energy efficiency, and it can be done in two types of networks i.e., homogenous and heterogeneous. WSNs having nodes of same energy level are called homogenous WSNs. Low-Energy Adaptive Clustering Hierarchy (LEACH) \cite{Reference 5}, Power Efficient Gathering in Sensor Information Systems (PEGASIS) \cite{Reference 7} and Hybrid Energy-Efficient Distributed Clustering (HEED) \cite{Reference 8} are examples of cluster based protocols which are designed for homogenous WSNs. These algorithms perform poor in heterogeneous WSNs. Nodes have less energy will expire faster than high energy nodes because these homogenous clustering based algorithms are incapable to treat every node with respect to energy. In heterogeneous WSNs, nodes are deployed with different initial energy. Stable Election Protocol (SEP) \cite{Reference 9}, Distributed Energy Efficient Clustering (DEEC) \cite{Reference 10}, Developed DEEC (DDEEC) \cite{Reference 11} and Enhanced DEEC (EDEEC) \cite{Reference 12} are examples of heterogenous WSN protocols.

\section{EDDEEC Protocol}
In this section, we present details of our EDDEEC protocol. Our proposed protocol implements the same idea of probabilities for CH selection based on initial, remaining energy level of the nodes and average energy of network as supposed in DEEC.

The average energy of $r^{th}$ round from \cite{Reference 10} is given as:

\begin{eqnarray}
\bar{E}(r)= \frac{1}{N}E_{total}(1-\frac{r}{R})
\end{eqnarray}

$R$ denotes total rounds during network lifetime and can be estimated from \cite{Reference 10} as:

\begin{eqnarray}
R= \frac{E_{total}}{E_{round}}
\end{eqnarray}

$E_{round}$ is the energy dissipated in a network during single round and calculated as:

\begin{eqnarray}
\begin{split}
E_{round}= L(2NE_{elec}+NE_{DA}+k\varepsilon_{mp}d_{to BS}^{4}+N\varepsilon_{fs}d_{to CH}^{2})
\end{split}
\end{eqnarray}

Where, $k$ is the number of clusters, $E_{DA}$ is the data aggregation cost expended in CH, $d_{toBS}$ is the average distance between CH to BS and $d_{toCH}$ is the average distance between cluster members to CH.

Now $d_{to BS}$ and $d_{to CH}$ can be calculated as:

\begin{eqnarray}
d_{to CH}= \frac{M}{\sqrt{2 \pi k}}, d_{to BS}= 0.765\frac{M}{2}
\end{eqnarray}

Through finding the derivative of $E_{Round}$ with respect to, $k$ to zero, we get the $k_{opt}$ optimal number clusters as:

\begin{eqnarray}
k_{opt}=\frac{\sqrt{N}}{\sqrt{2\pi}}\sqrt{\frac{\varepsilon_{fs}}{\varepsilon_{mp}}}\frac{M}{d_{toBS}^{2}}
\end{eqnarray}

At start of each round, node decides whether to become a CH or not based on threshold calculated by the following equation and as supposed in \cite{Reference 5, Reference 10}.

\begin{eqnarray}
T(s_{i})=
\begin{cases}
\frac{p_{i}}{1-p_{i}(rmod\frac{1}{P_{i}})} & if\; s_{i}\;\epsilon \;G \\
0 & otherwise
\end{cases}
\end{eqnarray}

where, $G$ is the set of nodes eligible to become CH for round $r$ and $p$ is the desired percentage of CH.
In real scenarios, WSNs have more than two types of heterogeneity. Therefore, in EDDEEC, we use concept of three levels heterogeneity and characterized the nodes as normal, advance and super nodes as supposed in \cite{Reference 12}. The probability for three types of nodes given by EDEEC is given below:

\begin{eqnarray}
p_{i}=
\begin{cases}
\frac{p_{opt}E_{i}(r)}{(1+m(a+m_{o}b))\bar{E}(r)}   & if \; s_{i} \;is\; the\; normal \;node\\
\frac{p_{opt}(1+a)E_{i}(r)}{(1+m(a+m_{o}b))\bar{E}(r)}  & if \;s_{i}\; is \;the \;advanced \;node\\
\frac{p_{opt}(1+b)E_{i}(r)}{(1+m(a+m_{o}b))\bar{E}(r)}  & if \;s_{i} \; is \;the\; super \;node\\
\end{cases}
\end{eqnarray}

The difference between DEEC, DDEEC, EDEEC and EDDEEC is generalized in Eq. 7, which defines probabilities to become CH for current round. Aim of this expression is to distribute energy consumption over network efficiently, increase stability period and lifetime of network. However, after some rounds, some super and advance nodes have same residual energy level as normal nodes due to repeatedly CH selection. Although EDEEC continues to punish advance and super nodes. Same is the problem with DEEC, it continues to punish just advance nodes and DDEEC is only effective for two-level heterogenous network as mentioned previously in related work. To avoid this unbalanced case in three-level heterogenous network and to save super and advance nodes from over penalized, we propose changes in function which defined by EDEEC for calculating probabilities of normal, advance and super nodes. These changes are based on absolute residual energy level $T_{absolute}$, which is the value in which advance and super nodes have same energy level as that of normal nodes. The idea specifies that under $T_{absolute}$ all normal, advance and super nodes have same probability for CH selection. Our proposed probabilities for CH selection in EDDEEC are given as follows:

\begin{eqnarray}
p_{i}=
\begin{cases}
\begin{split}
\frac{p_{opt}E_{i}(r)}{(1+m(a+m_{o}b))\bar{E}(r)}   & for\; N_{ml} \;nodes\;\\& if\; E_{i}(r)>T_{absolute}\\
\frac{p_{opt}(1+a)E_{i}(r)}{(1+m(a+m_{o}b))\bar{E}(r)}  & for\; Adv \;nodes\;\\& if\; E_{i}(r)>T_{absolute}\\
\frac{p_{opt}(1+b)E_{i}(r)}{(1+m(a+m_{o}b))\bar{E}(r)}  & for\; Sup\; nodes\;\\& if\; E_{i}(r)>T_{absolute}\\
c\frac{p_{opt}(1+b)E_{i}(r)}{(1+m(a+m_{o}b))\bar{E}(r)} & for\; N_{ml},\; Adv,\; Sup\; nodes\;\\& if \;
 E_{i}(r)\leq T_{absolute}\\
\end{split}
\end{cases}
\end{eqnarray}
\normalsize

The value of absolute residual energy level, $T_{absolute}$, is written as:

\begin{eqnarray}
T_{absolute}= z E_{0}
\end{eqnarray}

where, $z \epsilon ({0,1})$. If $z=0$ then we have traditional EDEEC. In reality, advanced and super nodes may have not been a CH in rounds $r$, it is also probable that some of them become CH and same is the case with the normal nodes. So, exact value of $z$ is not sure. However, through numerous of simulations using random topologies, we try to estimate the closest value of $z$ by varying it for best result based on first dead node in the network and find best result for $z=0.7$. Therefore, $T_{absolute}= (0.7) E_{0}$.

\section{Simulations and Results}
In this section, we present simulation result for DEEC, DDEEC, EDEEC and EDDEEC for three-level and multi-level heterogeneous WSNs using MATLAB. WSNs consist of $N=100$ nodes which are randomly placed in a field of dimension $100m\times100m$. For simplicity, we consider all nodes are either fixed or micro-mobile and ignore energy loss due to signal collision and interference between signals of different nodes that are due to dynamic random channel conditions. In this scenario, we are considering BS is placed at center of network field.

The performance metrics use for evaluation of clustering protocols for heterogeneous WSNs are stability period, lifetime of the heterogeneous WSNs and data packets which are successfully sent to BS. In heterogeneous WSNs, we used radio parameters as mentioned in Table 1 for different protocols deployed in WSNs and estimated performance for the case of three-level and multi-level heterogeneous WSNs

\begin{table}[!h]
\caption{Simulation Parameters}
\begin{center}
    \begin{tabular}{ | p{2.5cm} | p{2.5cm} |}
    \hline
    Parameters &	Values\\ \hline
    Network field & 100 m,100 m\\ \hline
    Number of nodes	& 100\\ \hline
    $E_{0}$(initial energy of normal nodes) & 0.5J\\ \hline
    Message size &	4000 bits \\ \hline
    $E_{elec}$ & 50nJ/bit\\ \hline
    $E_{fs}$ & 10nJ/bit/m2\\ \hline
    $E_{amp}$ & 0.0013pJ/bit/m4\\ \hline
    $EDA$ & 5nJ/bit/signal\\ \hline
    $d_{o}$(threshold distance) & 70m\\ \hline
    $P_{opt}$ & 0.1 \\ \hline
    \end{tabular}
\end{center}
\end{table}

We consider a network containing 20 normal nodes having $E_{0}$ energy, 32 advanced nodes having 2.0 times greater energy as compare to normal nodes and 48 super nodes containing 3.5 times greater energy as compare to normal nodes. Fig. 1 depict number of alive nodes during lifetime of network. First node for DEEC, DDEEC, EDEEC and EDDEEC dies at 969, 1355, 1432 and 1717 rounds, respectively, and all nodes dies at 5536, 5673, 8638 and 8638 rounds respectively. Fig. 2 shows that data sent to BS is more for EDDEEC than rest of the chosen protocols. Results show that EDDEEC is most efficient among all protocols in terms of stability period, network life time and packets sent to BS even in case of network containing more super and advanced nodes as compared to the normal nodes.

The parameters use for the simulation are given following:

\begin{itemize}
\item$m=0.8$
\item$m_{0}=0.6$
\item$a=2.0$
\item$b=3.5$
\end{itemize}

\begin{figure}[!h]
\centering
\includegraphics[height=5cm, width=8.5cm]{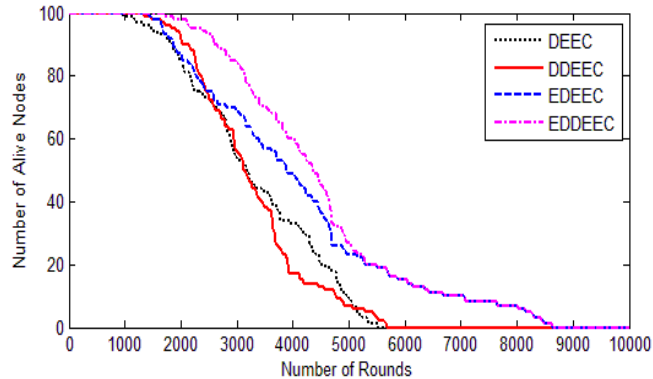}
\caption{Alive Nodes During Network Lifetime}
\end{figure}

\begin{figure}[!h]
\centering
\includegraphics[height=5cm, width=8.5cm]{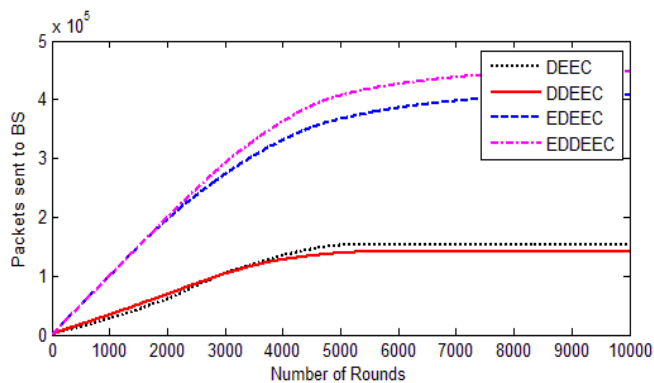}
\caption{Packets send to Base Station}
\end{figure}

\section{Conclusion}
In this paper, EDDEEC protocol is proposed for WSNs. EDDEEC is adaptive energy aware protocol which dynamically changes the probability of nodes to become a CH in a balanced and efficient way to distribute equal amount of energy between sensor nodes. We perform extensive simulations to check the efficiency of newly proposed protocol. The selected performance metrics for this analysis are stability period, network lifetime and packets sent to BS. The simulation analysis showed batter results which differentiate EDDEEC more efficient and reliable than DEEC, DDEEC and EDEEC.

\bibliographystyle{plain}

\end{document}